\documentclass[useAMS,usenatbib]{mn2e}
\usepackage{graphicx}
\usepackage{txfonts}
\usepackage{natbib}

\title[X-ray timing and spectral analysis of NGC~5408 X--1]{X-ray variability and energy spectra from NGC5408 X-1 with {\it XMM-Newton} }

\author[Caballero-Garc\'{i}a, Belloni \& Wolter]{M. D. Caballero-Garc\'{i}a$^{1}$\thanks{E-mail:
mcaballe@brera.inaf.it}, T.~M. Belloni$^{1}$, A. Wolter$^{2}$\\
$^{1}$ INAF-Osservatorio Astronomico di Brera, Via E. Bianchi 46, I-23807 Merate (LC), Italy \\
$^{3}$ INAF, Osservatorio Astronomico di Brera, via Brera 28, 20121 Milano, Italy
}

\begin{document}


\pagerange{\pageref{firstpage}--\pageref{lastpage}} \pubyear{2002}

\maketitle

\label{firstpage}

\begin{abstract}
The notion of source states characterizing the X-ray emission from black hole binaries has revealed to be a very
useful tool to disentangle the complex spectral and aperiodic phenomenology displayed by those classes of
accreting objects. We seek to use the same tools for Ultra-Luminous X-ray (ULX) sources.
We analyzed the data from the longest observations obtained from the ULX source in NGC~5408 (NGC~5408 X--1) taken by
{\it XMM-Newton}. We performed a study of the timing and spectral properties of the source.
In accordance with previous studies on similar sources, the intrinsic energy spectra of the source are well described by a cold accretion disc emission
plus a curved high-energy emission component. We studied the broad-band noise variability
of the source and found an anti-correlation between the root mean square variability in the 0.0001--0.2\,Hz and intensity,
similarly to what is observed in black-hole binaries during the hard states. We discuss the physical processes responsible for the
X-ray features observed and suggest that NGC~5408 X--1 harbors a black hole accreting in an unusual bright hard-intermediate state.

\end{abstract}


\section{Introduction} \label{introd}

Ultra-Luminous X-ray sources (ULXs) are point-like, off-nuclear, extra-galactic sources, with observed X-ray luminosities
(${\rm L}_{\rm X}{\ge}10^{39}\,{\rm erg}\,{\rm s}^{-1}$) higher than the Eddington luminosity for a stellar-mass black-hole
(${\rm L}_{\rm X}{\approx}10^{38}\,{\rm erg}\,{\rm s}^{-1}$). The true nature of these objects is still debated
\citep{feng11,fender12} as there is still no unambiguous estimate for the mass of the compact object hosted in these systems.
Assuming an isotropic emission, in order to avoid the violation of the Eddington limit, ULXs
might be powered by accretion onto Intermediate Mass Black Holes (IMBHs) with masses in the range
$10^{2}-10^{5}\,{\rm M}_{\odot}$ \citep{colbert99}. 
It has been also suggested that ULXs appear very luminous due to a
combination of moderately high mass, mild beaming and mild super-Eddington emission and that ULXs are
an inhomogeneous population composed of more than one class \citep{colbert99,fabbiano06}. 

ULXs have been studied intensively over the last decades (see \citealt{feng11} for a review). As in the case of Black Hole
Binaries (BHBs), some ULXs undergo spectral transitions from a {\it power-law state} state to a {\it high-soft state}
(see \citealt{belloni10b} and reference therein for a description of the spectral states in BHBs). Nevertheless, the classification of ULXs into 
canonical BH states is much more uncertain than in BHBs \citep{makishima07,soria11}. During the {\it power-law state} the
spectra of ULXs show a power-law spectral shape in the 3-8\,keV spectral range, together with a high-energy
turn-over at 6-7\,keV, and a {\it soft excess} at low energies (e.g. \citealt{kaaret06}). This {\it soft excess} can be modelled
by emission coming from the inner accretion disc and is characterized by a low inner disc temperature of ${\approx}0.2$\,keV. This 
is expected if the black holes in these sources are indeed IMBHs \citep{miller03,miller04}. Other explanations for the {\it soft excess}
imply a much smaller mass for the black hole in these sources, based on the idea that the accretion in the disc is not intrinsically standard, 
in contrast to the majority of BHBs (e.g. see \citealt{kajava09}). \citet{soria07} and \citet{mapelli09} suggest that at least some ULXs are 
consistent with black holes accreting at moderate rate with masses of ${\approx}50-100\,{\rm M}_{\odot}$,
as also supported by the analysis of N10 in the Cartwheel galaxy \citep{pizzolato10}.

The study of the timing properties of ULXs represent a promising way to confirm the associations/similarities with this class of sources and BHBs.
Variability in accreting stellar-mass black holes is usually studied by means of the Fourier Analysis, which allows to produce Power Density 
Spectra (PDS) where several different components are usually observed. BHBs PDS are usually composed of broad components
(in the form of red noise or band-limited noise) and narrow components (called Quasi Periodic Oscillations, QPOs). Although
we still know little about the physical origin of these temporal features, their phenomenology has proved useful in defining and identifying 
accretion states \citep{belloni10b}. \citet{heil09} performed a study of the fast time variability on various ULXs and found that for some sources 
the fast variability is suppressed for a reason which is currently unknown. \citet{middleton11} proposed
that the fast variability is suppressed for the sources with small inclination angle where the wind does not enter the line-of-sight. 
\citet{heil09} also found that a group of ULXs (including NGC~5408 X--1) have similar variability and PDS as
luminous BHBs and Active Galactic Nuclei (AGN) in the observed frequency band-pass ($10^{-3}-1$\,Hz). 

In this work we apply the Root Mean Square (rms)-Intensity Diagram (\citealt{munoz11}), which
has been proved to be useful to map states in BHBs (without the need of any spectral information). We perform timing and spectral studies, focusing 
on the evolution of the broad-band noise and on its dependence on the spectral properties of the source.

\subsection{NGC~5408 X--1}

The ULX in NGC~5408 X--1 was discovered with the {\it Einstein} observatory (Stewart et~al., 1983) and its {\it soft excess} found with {\it ROSAT}
\citep{fabian93}. It is located in a close-by ($D=4.8$\,Mpc, \citealt{karachentsev02}) small (size of $2.2{\times}1.1$\,kpc) starburst galaxy (\citealt{soria06} and references therein) 
and at ${\approx}20$\,arcsec from the centre of the galaxy. This ULX 
peaks in X-ray luminosity above $L_{X}=1{\times}10^{40}$\,${\rm erg\,s^{-1}}$ and is relatively nearby. \citet{grise12} identified the optical counterpart possibly as a B0I supergiant star.
Spherically-symmetric nebulae around NGC~5408 X--1 have been detected in radio and optical bands \citep{pakull03,soria06,lang07,cseh12}. This might indicate the presence of
strong winds from a high-mass accretion rate source. \citet{strohmayer09}
found a QPO in its PDS centred at $0.01$\,Hz and inferred a mass for the black hole in the range of $10^{3}-10^{4}\,{\rm M}_{\odot}$. Nevertheless, \citet{middleton11} propose
a much smaller mass ($10^2\,{\rm M}_{\odot}$) in base of the QPO and the timing properties. They propose that NGC~5408 X--1 is accreting in a super-Eddington regime and that
the QPO is analogous to the ultra-Low-Frequency QPO seen occasionally in a few BHBs. Recently, \citet{dheeraj12} studied the timing and
spectral properties of NGC~5408 X--1 and have found that the QPO frequency is variable (within the frequency range 0.0001-0.19\,Hz) and largely independent on the spectral
parameters. They suggested that NGC~5408 X--1 is accreting in the {\it saturation regime} (increase of the QPO frequency with constant disc flux and power-law photon index) frequently observed in BHBs \citep{vignarca03}.

\section{Observations and data reduction} \label{observ}

In this work we considered the 6 long (120-130\,ks) high-quality observations of NGC~5408 X--1 collected by the {\it XMM-Newton} satellite over 6 years (2006-2011). 
The EPIC camera was operating in the {\it Full Frame} mode and with the {\it Thin Filter} set (see Tab.~\ref{table_obs}). Some of the observations were affected by relatively
high background rates (flaring). We removed these periods for the timing data analysis by inspecting the light curve of the source. For the spectral analysis, 
we applied the standard filtering of removing time periods with count-rates in the FOV higher than $0.4\,{\rm cts}/{\rm s}$ (EPIC-pn only).

For the timing analysis, we filtered the EPIC pn+MOS event files, selecting only the best-calibrated events (pattern${\le}4,12$ for the pn and
the MOS, respectively), and rejecting flagged events (i.e. keeping only flag$=0$ events) from a circular region on the source (centre at coordinates ${\rm RA}=14\,h03\,m19.6\,s$,
${\rm Dec}=-41\,d22\,m59.6\,s$; \citealt{gladstone09}) and radius $28\,$arcsec. The radius chosen contains ${\approx}90\%$ of the source photons, which is an optimal choice for not including too much
background \footnote{See Fig~7 from the {\it XMM-Newton Users Handbook}: \\
http://xmm.esac.esa.int/external/xmm\_user\_support/documentation/ \\
uhb/XMM\_UHB.pdf}. No background subtraction was applied since the background contribution was found to be small.
We payed particular attention to extract the list of photons not randomized in time. To this purpose, we used the tasks {\it epchain} for the pn and  
{\it emproc} (with {\it randomizetime=no}) for the MOS cameras, respectively. 

\begin{figure*}
\centering
 \includegraphics[bb=36 270 577 522,width=12.0cm,angle=0,clip]{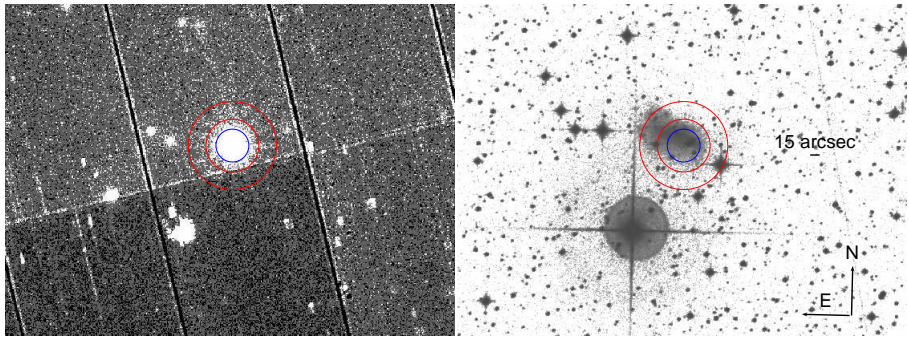}
 \caption{($12{\times}12$\,arcmin) FOV of NGC~5408 X--1 with {\it XMM-Newton}/EPIC-pn (flag$=0$ events; left) and (optical) STSCI-DSS I/II image (right). Regions for the spectral extraction of events of the source (circle/blue) and the background (annulus/red) are shown.}
 \label{plot_region}
\end{figure*}

For the spectral analysis we used only the EPIC pn camera, in order to avoid issues due to cross-calibration effects. Additionally, the EPIC pn camera
has a higher effective area (i.e. double) than each one of the MOS cameras and has sufficient statistics for the spectral fitting. 
We filtered the event files, selecting only the best-calibrated events (pattern${\le}4$ for the pn), and rejecting flagged events (flag$=0$).
We extracted the flux from a circular region on the source centred at the coordinates of the source 
and radius $28\,$arcsec. The background was extracted from 
an annular region (with inner and outer radius of $45,75\,$arcsec, respectively) centred at the coordinates of 
the source (see Fig.~\ref{plot_region}). To check the dependence of our results on these choices we also performed a background subtraction from a region in the same chip of the 
source and far away from it and from the boundaries and have seen no difference in the results shown in this paper.  
We built response functions with the {\it Science Analysis System} (SAS) tasks {\tt rmfgen} and {\tt arfgen}. We 
fitted the background-subtracted spectra with standard spectral models
in XSPEC 12.7.0 \citep{a1}. All errors quoted in this work are $68\%$ ($1{\sigma}$) confidence. The spectral fits were limited to the 0.3-10\,keV range, where the 
calibration of the instruments is the best. The spectra were rebinned in order to have at least 25 counts for each background-subtracted spectral channel 
in order to perform the chi-squared fitting and to avoid oversampling of the 
intrinsic energy resolution by a factor larger than 3 \footnote{As recommended in: \\
http://xmm.esac.esa.int/sas/current/documentation/threads/PN\_spectrum\_thread.shtml}.

\section{Analysis}  \label{timing}

\subsection{Timing Analysis}  \label{timing}

We performed an analysis of the fast time variability of NGC~5408 X--1 separately in the 0.3-1\,keV and 1-10\,keV energy ranges. The (0.3-10\,keV) EPIC count
rate is $(1.2-1.6)\,{\rm cts}\,{\rm s}^{-1}$ on average. The time resolution of the instruments is 2.6\,s (MOS) and 73.4\,ms (pn).

For both analysis we took into account good time intervals (GTIs) similar to the ones by \citet{dheeraj12} (see their Fig.~1). We excluded
the periods of high-background with the introduction of a minimum 
number of gaps in the light curve that might produce artificial noise in our study of the intrinsic variability from the source.  
These time periods (gaps) were excluded from the PDS generation. We used the GHATS package, developed under the IDL environment 
at INAF-OAB \footnote{http://www.brera.inaf.it/utenti/belloni/GHATS\_Package/Home.html}, 
to produce the PDS from 2048 points in each light curve. The PDS 
were then averaged together for each observation. The pn+MOS light curve was binned at the lowest time resolution of the two (2.6\,s). This yields a Nyquist frequency of $=0.19$\,Hz. 
The PDS were normalized according to \citet{leahy83}. All of the PDS show low-frequency flat-topped noise. 

PDS fitting was carried out with the standard XSPEC fitting package by using a unit response. Fitting
the (0.3-1\,keV) PDS with a model constituted by a zero centred Lorentzian for the flat-topped noise plus a constant for the Poissonian noise results in
mostly acceptable chi-square values only for the PDS of Obs.~4,~6 (see Tab.~\ref{table_timing}). In the case of Obs.~1,~2,~3 the fit statistics with this 
model is worse, i.e. ${\chi}^{2}/{\nu}=134/98,120/98,101/98$, respectively. These observations and Obs.~5 (with a fit statistics of ${\chi}^{2}/{\nu}=88/98$ using the previous model) show positive residuals
at ${\approx}(0.002-0.02)$\,Hz (see Tab.~\ref{table_timing}) that we fitted by adding a further Lorentzian component centred at those frequencies. This component changed the fits statistics by 
${\Delta}{\chi}^{2}{\le}20$ for ${\Delta}{\nu}=3$ d.o.f., thus a ${\le}3{\sigma}$ improvement. Although not significant for Obs.~2,~3,~5 we took into account these features, since they
are broad and therefore affect substantially the measure of the rms, i.e. by $(1-3)\%$. 

The fit of the (1-10\,keV) PDS with a model constituted by a zero centred Lorentzian for the flat-topped noise plus a constant for the Poissonian noise results
in a poor description of the data. The fit statistics with this model is of ${\chi}^{2}/{\nu}=125/97,116/98,107/98,130/98,135/98,149/98$ for Obs.~1-6, respectively. In
all the observations there are positive residuals centred in the range of $(0.01-0.04)$\,Hz that we fitted by adding a further Lorentzian component centred at those frequencies.
This component changed the fits statistics by ${\Delta}{\chi}^{2}{\approx}33,10,16,20,50,33.5$ for ${\Delta}{\nu}=3$ d.o.f., thus a $(4.9,2.9,3.2,3.9,8.0,4.2){\sigma}$ improvement for 
Obs.~1-6, respectively. This is a significant improvement for all the cases. We took into account this feature, since it
is broad and therefore affects substantially our the measure of the rms, i.e. by ${\approx}(1-4)\%$ for all the observations. In the case of Obs.~1,~4,~6 there is an excess at low frequencies and fitting it 
with a power-law yields a non-significant component ($<3{\sigma}$) for Obs. 1,6 that again we took into account, since the change in the fractional rms is of the order of $(0.3-0.7)\%$. The photon index was
unconstrained, so we fixed its value to 2, which is the asymptotic high-frequency slope of a Lorentzian. Additionally, the (1-10\,keV) PDS of Obs.~2 shows positive residuals in the
form of a peak centred at ${\nu}_{\rm QPO}{\approx}0.01\,{\rm Hz}$, consistent with previous findings \citep{strohmayer09}. The detection of this peak is not significant ($2.3{\sigma}$, calculated
as the value of the Lorentzian normalization divided by its $1{\sigma}$ error). We did not include this component in our fits since it does not affect substantially the measure of the fractional rms ($<0.4\%$). 

\begin{figure*}
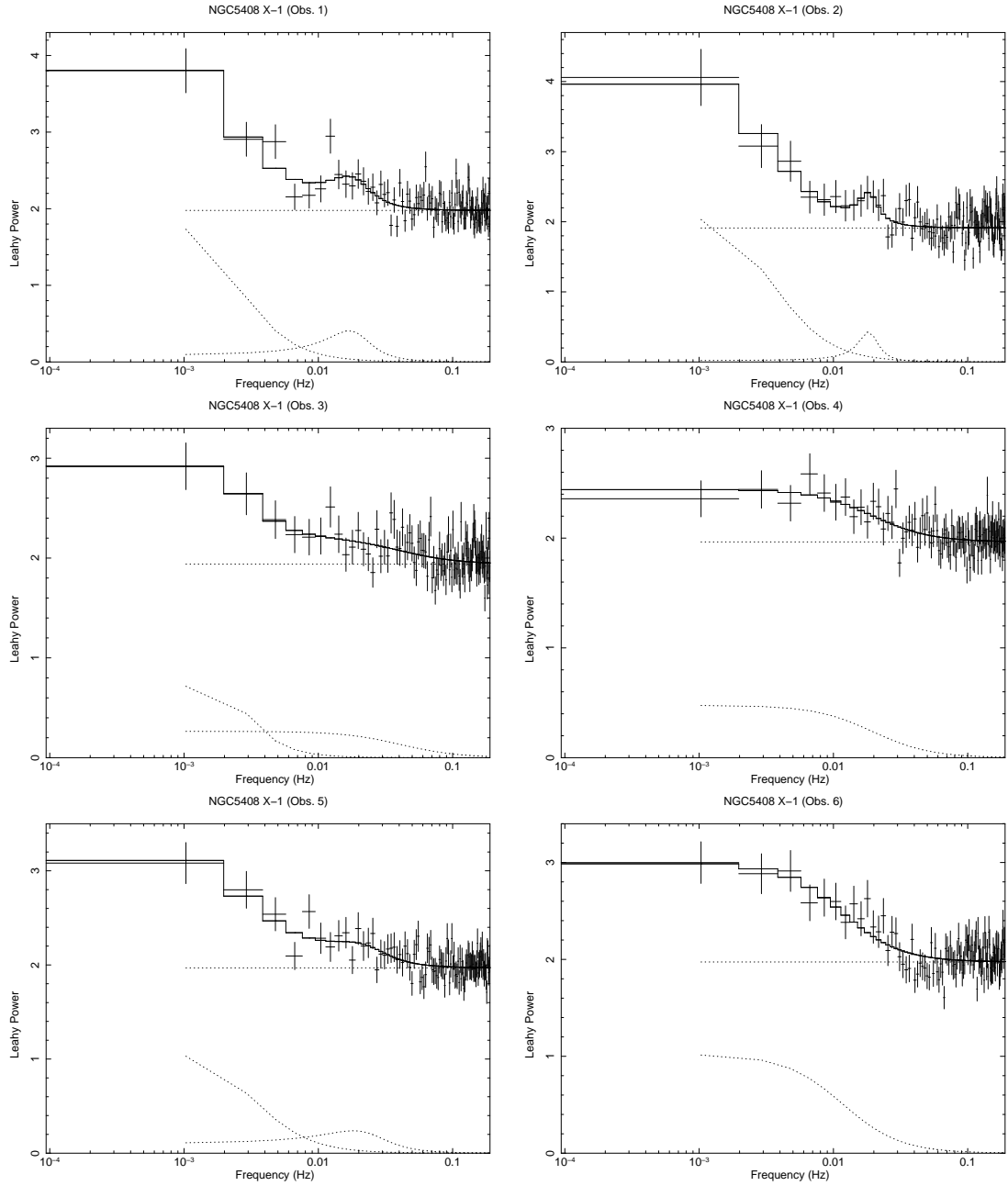

\centering
 \includegraphics[bb=34 5 570 700,width=5.7cm,angle=270,clip]{spds1.ps}
 \includegraphics[bb=34 5 570 700,width=5.7cm,angle=270,clip]{spds2.ps} 
 \includegraphics[bb=34 5 570 700,width=5.7cm,angle=270,clip]{spds3.ps}
 \includegraphics[bb=34 5 570 700,width=5.7cm,angle=270,clip]{spds4.ps}
 \includegraphics[bb=34 5 570 700,width=5.7cm,angle=270,clip]{spds5.ps} 
 \includegraphics[bb=34 5 570 700,width=5.7cm,angle=270,clip]{spds6.ps}
 \caption{Power density spectra of the {\it XMM-Newton} EPIC/pn+MOS data in the energy and frequency range (0.3-1\,keV) and 0.0001-0.19\,Hz, respectively, during observations 1--6 (top-left to bottom-right) with the best fit model (solid line) and the model components (dotted line). The results of the fitting are shown in Tab.~\ref{table_timing} and discussed in the text.}
 \label{plots_timing1}
\end{figure*}

\begin{figure*}
\centering
 \includegraphics[bb=34 5 570 700,width=5.7cm,angle=270,clip]{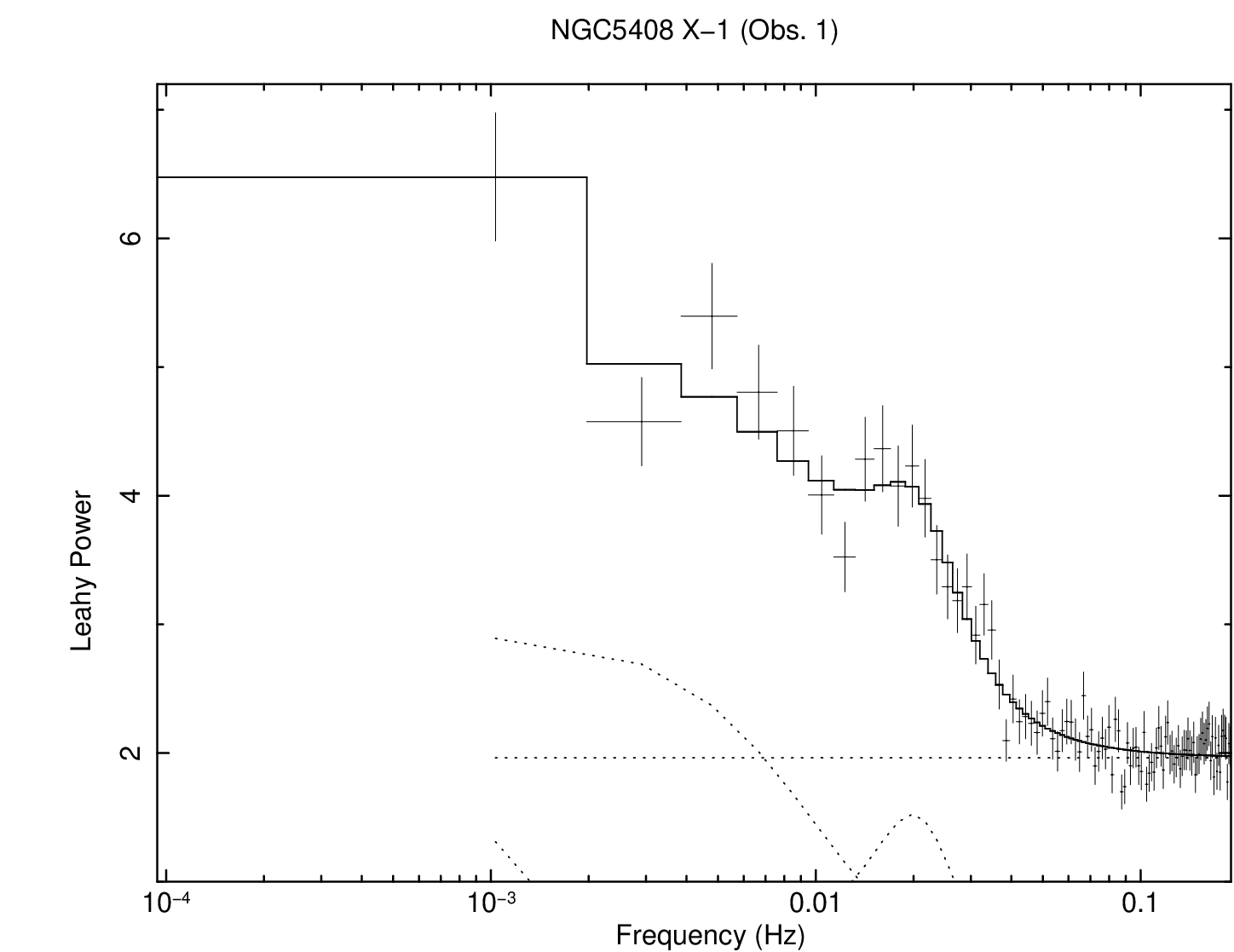}
 \includegraphics[bb=34 5 570 700,width=5.7cm,angle=270,clip]{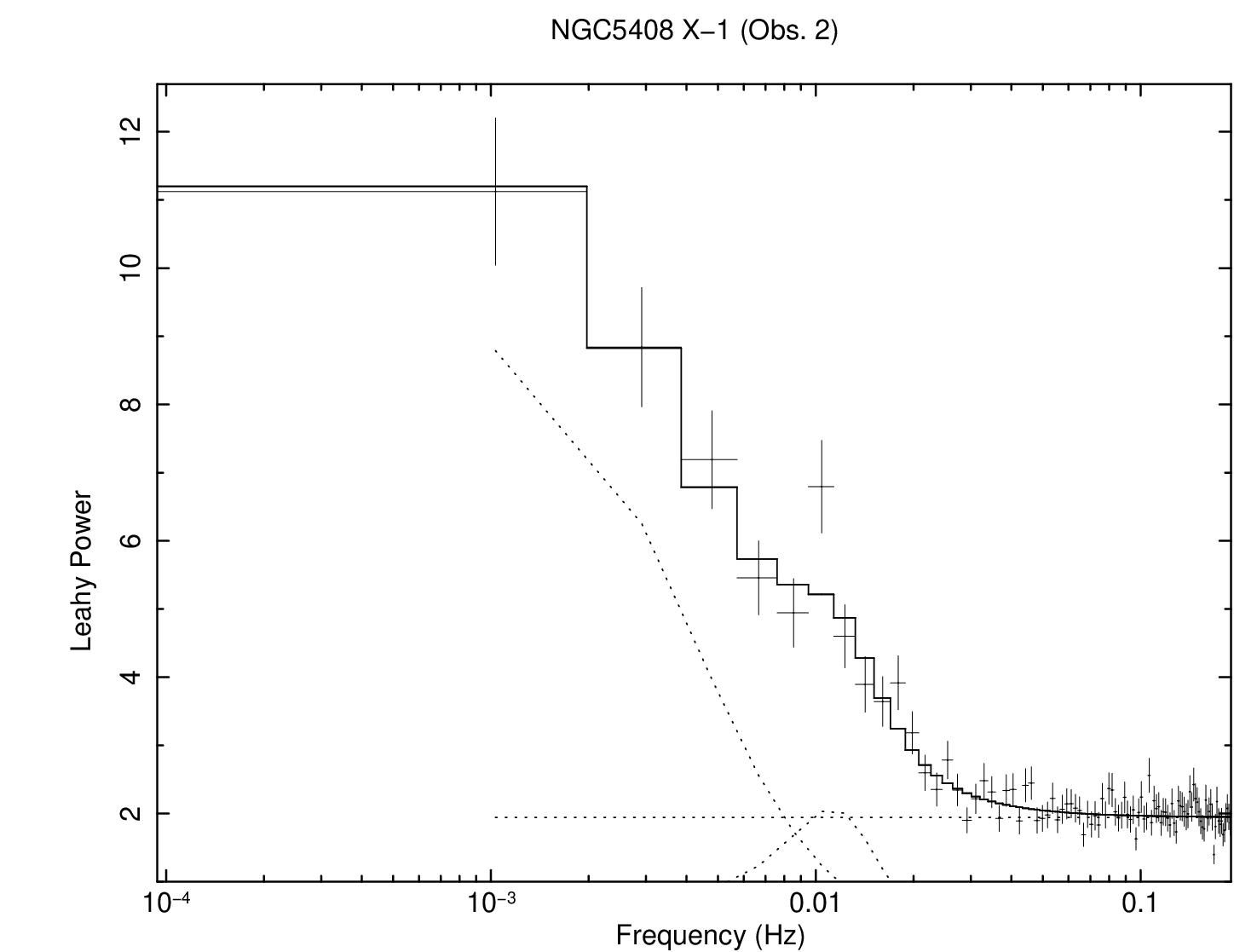}
 \includegraphics[bb=34 5 570 700,width=5.7cm,angle=270,clip]{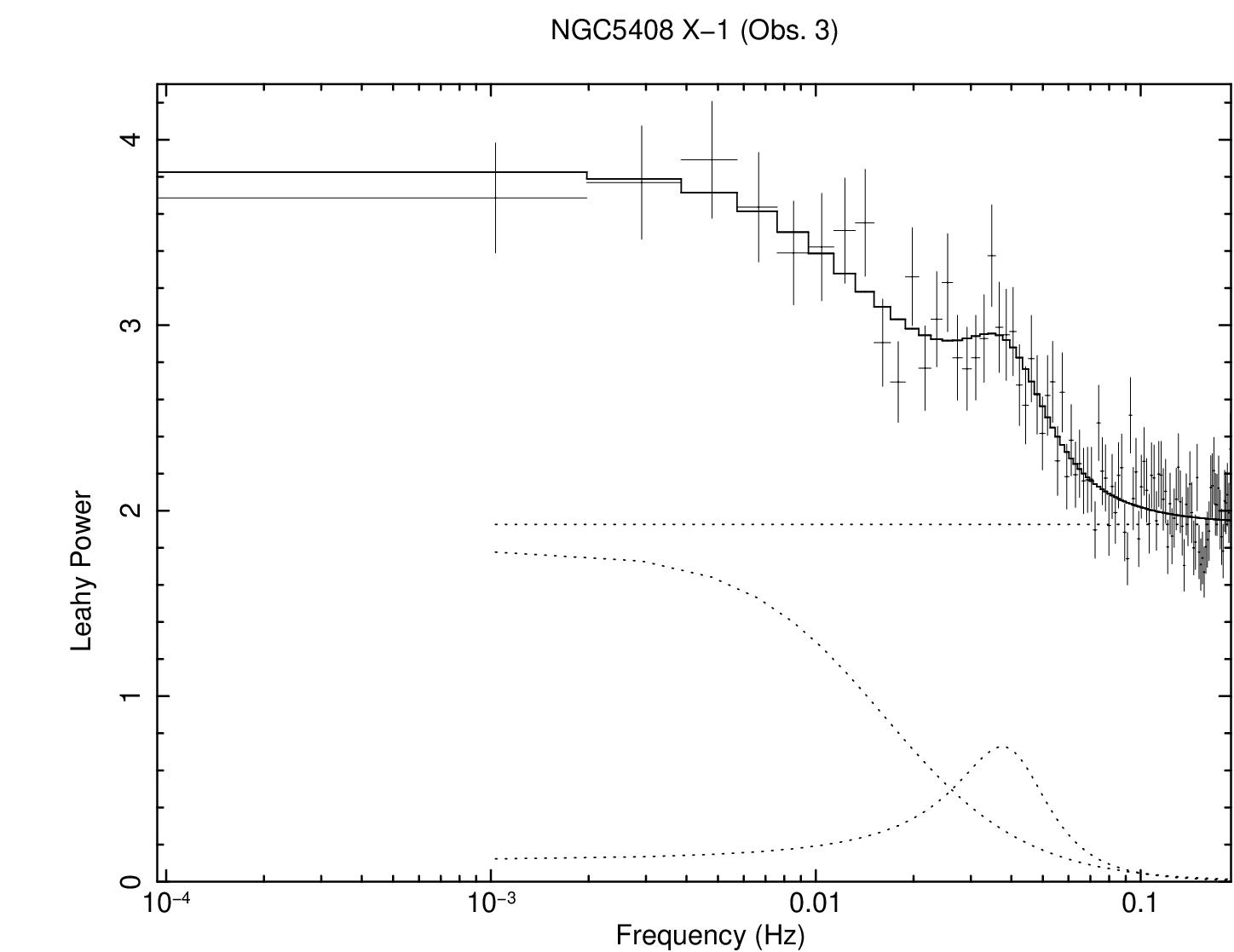}
 \includegraphics[bb=34 5 570 700,width=5.7cm,angle=270,clip]{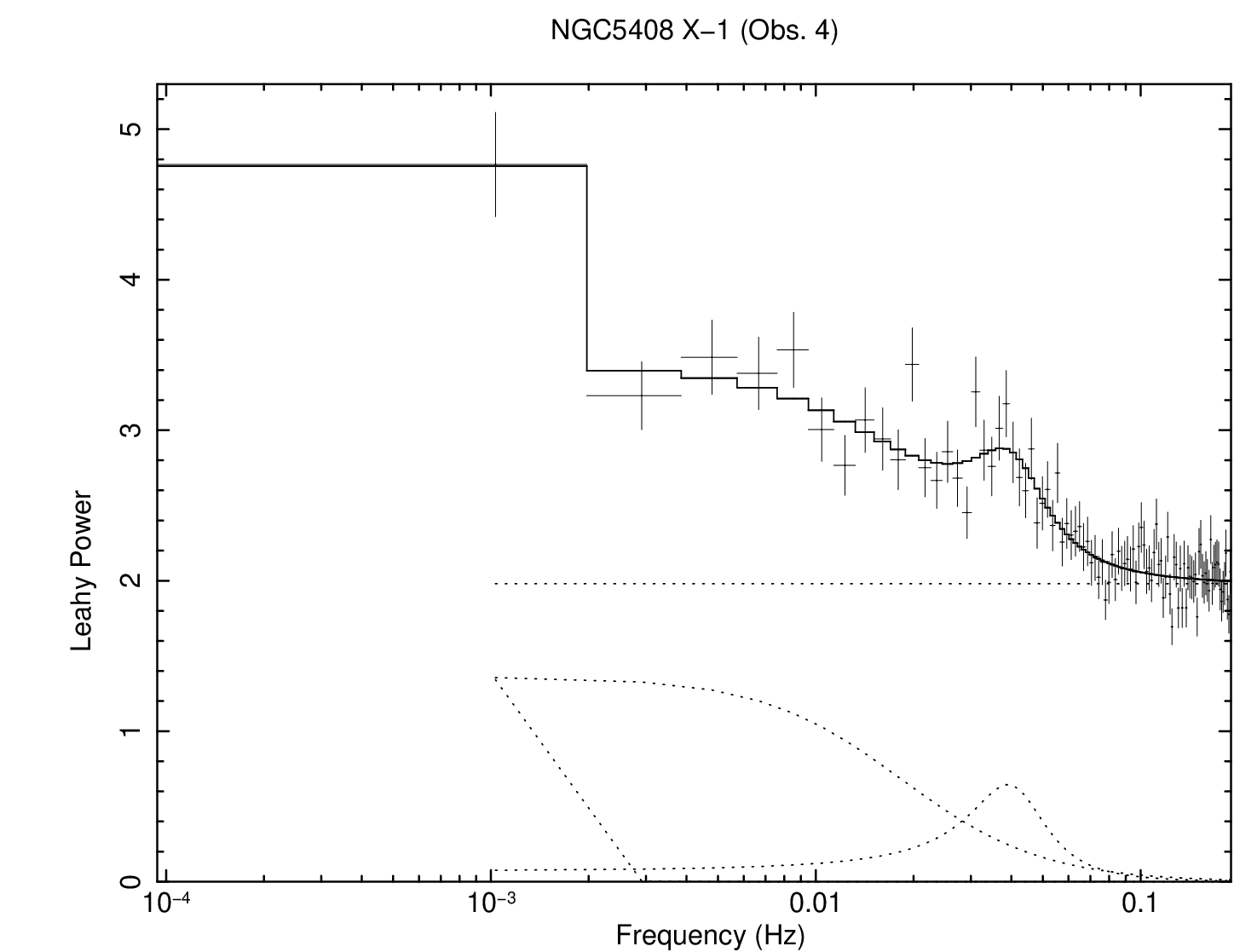}
 \includegraphics[bb=34 5 570 700,width=5.7cm,angle=270,clip]{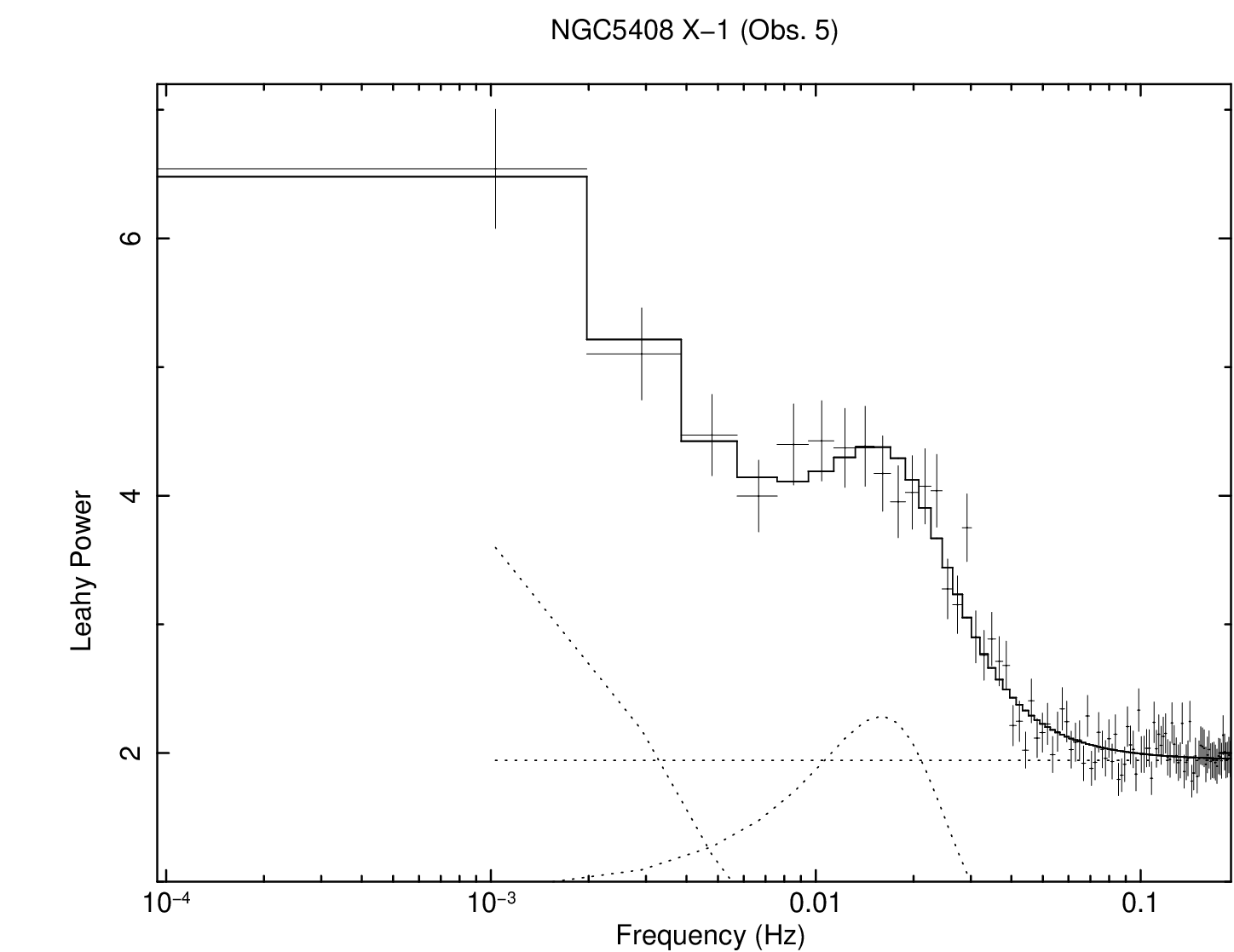}
 \includegraphics[bb=34 5 570 700,width=5.7cm,angle=270,clip]{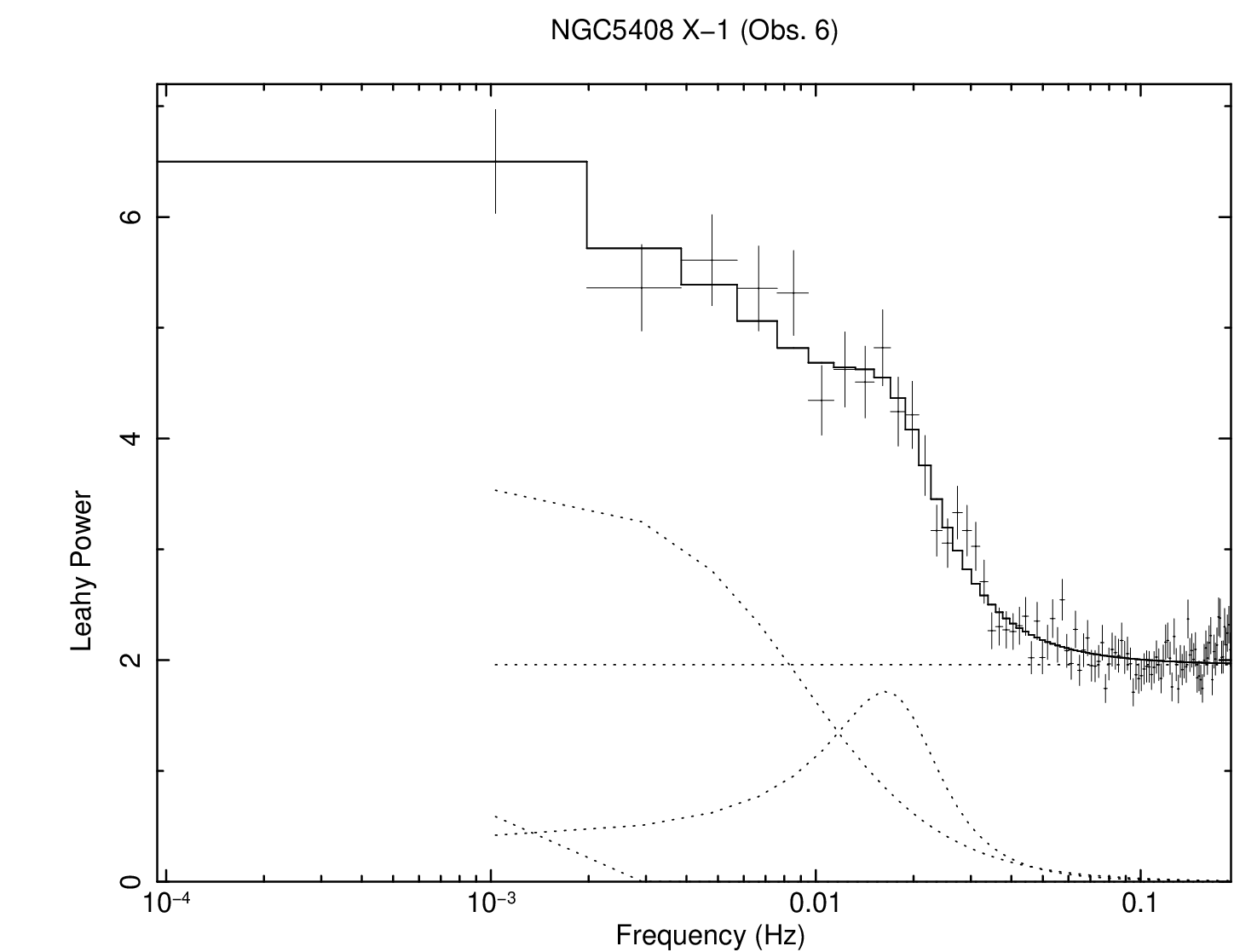}
 \caption{Power density spectra of the {\it XMM-Newton} EPIC/pn+MOS data in the energy and frequency range (1-10\,keV) and 0.0001-0.19\,Hz, respectively, during observations 1--6 (top-left to bottom-right) with the best fit model (solid line) and the model components (dotted line). The results of the fitting are shown in Tab.~\ref{table_timing} and discussed in the text.} 
 \label{plots_timing2}
\end{figure*}

We calculated the fractional rms from the best fit model (integrated in the 0.0001-0.19\,Hz band). This was found to be ${\approx}15\%$ and $(30-50)\%$ in the 0.3--1\,keV and 1--10\,keV energy ranges, respectively.
The values obtained are reported in Tab.\,\ref{table_timing}, together with the results from the fits and the count rates
in the different energy ranges. We plot in Fig. \ref{plots_timing1} and \ref{plots_timing2} the broad-band PDS with the best-fit model. We notice that 
the integration of the rms from the data itself,
without modeling, yields consistent values for the total fractional rms. 

\begin{table*}
 \centering
 \begin{minipage}{120mm}
  \caption{Log of the observations.}
  \label{table_obs}
  \begin{tabular}{@{}lcccccc@{}}
  \hline
                                &   Obs.~1                      &          Obs.~2        & Obs.~3                  &    Obs.~4             &    Obs.~5              &       Obs.~6     \\
 \hline
   Date (dd/mm/yyyy)                &   13/01/2006              &  13/01/2008            &   17/07/2010            &   19/07/2010          &  26/01/2011            &    28/01/2011                      \\
   Obs. ID                          &   0302900101              &   0500750101           &  0653380201             &  0653380301           &  0653380401            &    0653380501                      \\
   Exposure time\,(ks)\,$^{1}$      &   130            &  114                   &  127                    &  129                  &  119                   &    124                               \\
   Observation mode\,$^{2}$     &    Full Frame             &   Full Frame           &    Full Frame           &  Full Frame           &   Full Frame           &    Full Frame                        \\
\hline
\end{tabular}
\footnotetext{$^1$ The EPIC/pn exposure time.} 
\footnotetext{$^2$ From the EPIC cameras.} 
\end{minipage}
\end{table*}

\begin{table*}
 \centering
 \begin{minipage}{120mm}
  \caption{Results from the timing analysis.}
  \label{table_timing}
  \begin{tabular}{@{}lcccccc@{}}
  \hline
                                &   Obs.~1                      &          Obs.~2        & Obs.~3                  &    Obs.~4             &    Obs.~5              &       Obs.~6     \\
 \hline
                                &                               &                        & {\it Soft} (0.3-1\,keV) PDS                &                       &                        &                  \\
 \hline

   Count rate\,$^{3}$\,(${\rm cts}\,{\rm s}^{-1}$)   & $0.855{\pm}0.003$  &  $0.769{\pm}0.004$  &   $0.943{\pm}0.003$  &   $0.908{\pm}0.003$  &  $0.866{\pm}0.003$   &    $0.832{\pm}0.003$  \\
   ${\nu}_{\lor}(1)$\,(Hz)      &   $0$                     &  $0$                 &   $0$                   &   $0$                 &  $0$                   &     $0$              \\
   FWHM\,(Hz)                   &   $0.005{\pm}0.002$       &  $0.007{\pm}0.002$   &   $0.08{\pm}0.04$       &   $0.039{\pm}0.010$   &  $0.006{\pm}0.002$     &  $0.023{\pm}0.004$   \\
   Norm.                        &   $0.015{\pm}0.003$       &  $0.024{\pm}0.005$   &   $0.034{\pm}0.013$     &   $0.029{\pm}0.006$   &  $0.011{\pm}0.004$     &  $0.037{\pm}0.004$   \\
   ${\nu}_{\lor}(2)$\,(Hz)      &   $0.017{\pm}0.002$       &  $0.018{\pm}0.003$   &   $0.0013{\pm}0.0010$   &   --                  &  $0.018{\pm}0.011$     &  --                  \\
   FWHM\,(Hz)                   &   $0.018{\pm}0.005$       &  $0.007{\pm}0.005$   &   $0.004{\pm}0.003$     &   --                  &  $0.03{\pm}0.02$       &  --                  \\
   Norm.                        &   $0.011{\pm}0.003$       &  $0.005{\pm}0.002$   &   $0.004{\pm}0.003$     &   --                  &  $0.012{\pm}0.005$     &  --                  \\
   ${\Gamma}_{\rm poisson}$     &   $0$                     &  $0$                 &   $0$                   &   $0$                 &  $0$                   &  $0$                 \\
   ${\rm Norm.}_{\rm poisson}$  &   $1.98{\pm}0.02$         &  $1.91{\pm}0.02$     &   $1.94{\pm}0.04$       &   $1.97{\pm}0.02$     &  $1.97{\pm}0.02$       &  $1.97{\pm}0.02$      \\
   ${\chi}^{2}/{\nu}$           &   $111/95$                &   $112/95$           &  $94/95$                &   $70/98$             &  $77/95$               &   $107/98$           \\
   Fractional rms ($\%$)        &   $14.0{\pm}1.3$          &  $14.7{\pm}1.7$      &   $13.7{\pm}1.2$        &   $12.3{\pm}1.2$      &  $12.9{\pm}1.2$        &    $14.6{\pm}1.2$   \\
 \hline
                                &                               &                        & {\it Hard} (1-10\,keV) PDS                &                       &                        &                  \\
 \hline
   Count rate\,$^{3}$\,(${\rm cts}\,{\rm s}^{-1}$)   & $0.429{\pm}0.002$  &  $0.426{\pm}0.003$  &   $0.574{\pm}0.003$  &   $0.552{\pm}0.002$  &  $0.500{\pm}0.002$   &    $0.490{\pm}0.002$  \\
   ${\nu}_{\lor}(1)$\,(Hz)      &   $0$                     &  $0$                 &   $0$                   &   $0$                 &  $0$                   &  $0$                       \\
   FWHM\,(Hz)\,$^{2}$           &   $0.020{\pm}0.005$       &  $0.008{\pm}0.002$   &   $0.033{\pm}0.007$     &  $0.037{\pm}0.010$    &  $0.006{\pm}0.002$     &  $0.018{\pm}0.005$         \\
   Norm.                        &   $0.09{\pm}0.02$         &  $0.12{\pm}0.02$     &   $0.09{\pm}0.02$       &  $0.079{\pm}0.019$    &  $0.039{\pm}0.008$     &  $0.10{\pm}0.03$           \\
   ${\nu}_{\lor}(2)$\,(Hz)      &   $0.020{\pm}0.002$       &  $0.011{\pm}0.002$   &   $0.038{\pm}0.003$     &  $0.039{\pm}0.002$    &  $0.016{\pm}0.002$     &  $0.016{\pm}0.002$         \\
   FWHM\,(Hz)\,$^{2}$           &   $0.019{\pm}0.004$       &  $0.011{\pm}0.003$   &   $0.033{\pm}0.010$     &  $0.028{\pm}0.010$    &  $0.02{\pm}0.02$       &  $0.017{\pm}0.004$         \\
   Norm.                        &   $0.046{\pm}0.014$       &  $0.038{\pm}0.013$   &   $0.037{\pm}0.014$     &  $0.028{\pm}0.013$    &  $0.089{\pm}0.010$     &  $0.047{\pm}0.017$         \\
   ${\Gamma}$                   &   $2$                     &  --                  &   --                    &   $2$                 &  --                    &  $2$           \\
   Norm.                        &   $(2.4{\pm}1.1){\times}10^{-7}$   &  --         &   --                    &  $(2.5{\pm}0.7){\times}10^{-7}$       &  --            &  $(1.1{\pm}1.0){\times}10^{-7}$       \\
   ${\Gamma}_{\rm poisson}$     &   $0$                     &  $0$                 &   $0$                   &   $0$                 &  $0$                   &  $0$                 \\          
   ${\rm Norm.}_{\rm poisson}$  &   $1.96{\pm}0.02$         &  $1.94{\pm}0.02$     &   $1.93{\pm}0.03$       &   $1.98{\pm}0.02$     &  $1.94{\pm}0.02$       &  $1.96{\pm}0.02$      \\
   Fractional rms ($\%$)        &   $44.4{\pm}1.0$          &  $45.6{\pm}1.4$      &   $36.0{\pm}0.9$        &   $33.9{\pm}0.8$      &  $41.8{\pm}0.8$        &    $42.7{\pm}0.9$    \\
   ${\chi}^{2}/{\nu}$           &   $87/94$                 &   $101/95$           &  $91/95$                &   $93/94$             &  $78/95$               &   $114/94$        \\
\hline
\end{tabular}
\footnotetext{Values for the count rate and characteristics of the noise (using the model$^{1,2}$), separately in the 0.3-1\,keV (i.e. {\it Soft PDS}: top) and in the 1-10\,keV energy range (i.e. {\it Hard PDS}: bottom) for the six observations.}
\footnotetext{$^1$ Model used: A) {\tt lorentz+lorentz+powerlaw} for the 0.3-1\,keV energy range and B) {\tt lorentz+lorentz+powerlaw+powerlaw} for the 1-10\,keV energy range.  }
\footnotetext{$^2$ Errors are $68\%$ confidence errors.  }
\footnotetext{$^3$ Background-subtracted count rate from the pn+MOS cameras. }
\end{minipage}
\end{table*}

\subsection{Spectral Analysis}  \label{spectr}

Fig.~\ref{plot_region} shows the galaxy area 
in X-rays and in the optical. The regions of the extraction of the flux from the source (circle) and the background (annulus)
are also plotted. As can be seen, a small fraction of the inner region of the galaxy falls into the background extraction region. The (pn) count rate
from the (annulus) background region is $0.107{\pm}0.003\,{\rm cts}\,{\rm s}^{-1}$ in the (0.3-10\,keV) energy range.

We started fitting the spectra with an absorbed power-law model, using the Tuebingen-Boulder ISM absorption model ({\tt tbabs} in XSPEC) to account
for the interstellar absorption (${\rm N}_{\rm H}=7{\times}10^{20}\,{\rm cm}^{-2}$ in the direction to NGC~5408; \citealt{dickey90}). This parameter was set free to vary
in order to account for intrinsic absorption. We fitted the spectra simultaneously, constraining the column density to be the same for all the spectra. With 
this model we obtained a bad description
of the spectra ${\chi}^{2}/{\nu}{\gg}2$ (with ${\nu}=710$ d.o.f.), with positive residuals at ${\le}2$\,keV and high-energy curvature at ${\gtrsim}5$\,keV. The spectra are curved with a 
break at ${\approx}5-7$\,keV, in agreement with what has been found in previous studies (see e.g. \citealt{stobbart06,gladstone09,caballero10}).
To account for the low-energy positive residuals we added a model constituted by an absorbed multicolor inner emission disc ({\tt diskbb} or {\tt diskpn} in XSPEC).
The second is an extension of the first, including corrections for the temperature distribution near the black hole. To account for the high-energy residuals we added a cut-off, replacing the power-law component by 
an exponential rolloff ({\tt cutoffpl} in XSPEC) model to fit the high-energy spectra. This model improved the fit substantially (${\chi}^{2}/{\nu}{\approx}1.6$, with ${\nu}=692$).
We also tried by substituting the cutoff power-law by the more physical {\tt compTT} model the latter with the temperature for the input photons equal to the inner
disc temperature and obtained very similar quality of the fits (${\chi}^{2}/{\nu}{\approx}1.6$, with ${\nu}=692$). The resulting parameters are in agreement with previous
studies \citep{gladstone09,middleton11,dheeraj12}.

With the continuum model adopted (i.e. absorption, curved power-law and accretion disc) there are excesses at low
energies (around 0.6 and 1\,keV) that we attribute to the diffuse emission from the galaxy. To account for them we 
had to include two {\tt apec} models (one was not enough) with temperatures of ${\approx}1,0.1$\,keV. We obtained a good
fit (${\chi}^{2}/{\nu}{\approx}1.0$, with ${\nu}=688$ for all the spectral models used; see Tab.~\ref{table_spe}).
The parameters of the {\tt apec} models were constrained to be the same between the observations. We fixed the metal abundances to 
${\rm Z}=0.5\,{\rm Z}_{\odot}$ (sub-solar metallicities were found by 
\citealt{mendes06}). Nevertheless, the results do not significantly differ from the case of solar abundances and are insensitive of the sub-solar value adopted.
The maximum temperature found for the diffuse X-ray component is by far too high compared to what is expected in regular
non-starburst galaxies (i.e. ${\rm kT}_{\rm apec}=0.1-0.2$\,keV). Indeed, NGC~5408 is a galaxy with a central 
starburst region (see \citealt{soria06} and references therein). It has been seen (e.g. \citealt{buote98,warwick07}) that the highest X-ray
quality data from starburst galaxies show plasmas in a multi-phase state, with two thermal diffuse emission components in their spectra, one cold and one hot
(${\rm kT}_{\rm apec}=(0.1-0.6)\,{\rm keV},(0.8-1.0)$\,keV), the hottest associated with the central starburst, in agreement with our findings.

The most X-ray luminous diffuse components are present in starburst galaxies
(e.g. NGC~3256, NGC~253, M~82; \citealt{buote98,moran99}).
For example, the total inferred luminosity of the diffuse plasma from the inner region (within a radius of 10.5\,kpc in the plane
of the galaxy) of the spiral star-forming galaxy M~101 is
${\rm L}_{\rm X}(0.5-2\,{\rm keV})=2.1{\times}10^{39}$\,${\rm erg}{\rm s}^{-1}$ \citep{warwick07}.
The intrinsic (i.e. unabsorbed) luminosity of the diffuse plasma (i.e. from the {\tt apec} model described above)
from the inner (0.6\,kpc) region of NGC~5408 is
${\rm L}_{\rm X}(0.3-10\,{\rm keV}){\approx}1{\times}10^{39}$\,${\rm erg}\,{\rm s}^{-1}$ (assuming a distance of 4.8\,Mpc), thus ${\approx}10\%$ of the total 
(0.3-10\,keV) intrinsic luminosity from the ULX.
This value closely resembles the total luminosity of the diffuse thermal components of the dwarf starburst galaxy NGC~1569 
(${\rm L}_{\rm X}(0.3-6\,{\rm keV})=8.8{\times}10^{38}$\,${\rm erg}{\rm s}^{-1}$; \citealt{martin02}).

\begin{figure}
\centering
 \includegraphics[bb=41 -18 564 700,width=6.0cm,angle=270,clip]{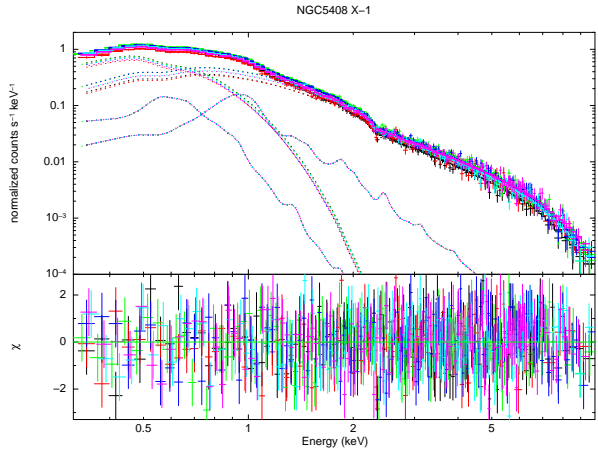}
 \caption{EPIC-pn {\it XMM-Newton} spectra (top) and chi-square residuals (bottom) of NGC5408 X--1 during observations 1--6 (in black, red, green, dark and light blue and magenta, respectively) fitted with the spectral model C. }
 \label{plot_spe}
\end{figure}

When using the {\tt compTT} model to describe the high-energy component of the spectra (models C and D in Tab. 3), the obtained temperature of the electrons 
in the corona was ${\rm kT}_{\rm e}{\le}5$\,keV and the optical depth ${\tau}=3-6$.
In the phenomenological model with {\tt cutoffpl} (model A and B in Tab. 3), the photon indices and the high-energy cut-off found are in the range $1.9-2.3$ and $4-10$\,keV, respectively.
The photon indices are very similar to those obtained from BHBs during the low/hard state
(${\Gamma}=1.7-2.0$, ${\rm kT}_{\rm in}{\approx}0.17$\,keV, \citealt{reis10}) but the high-energy cut-off is much lower than 
those observed in BHBs (${\rm E}=60-{\gtrsim}100$\,keV; \citealt{delsanto09}). The two high-energy spectral models provide the same statistical description to the data. 
The most relevant results of this spectral analysis and the derived unabsorbed fluxes are
in Tab.\,\ref{table_spe} and in Fig.~\ref{plot_spe}. The errors on the total flux (plus the flux from every individual component) were calculated using the {\tt cflux} component in XSPEC.

\begin{table*}
 \centering
 \begin{minipage}{160mm}
  \caption{Results from the spectral analysis.}
  \label{table_spe}
  \begin{tabular}{@{}lcccccc@{}}
  \hline
   Spectral parameter~$^1$$^,$$^2$ &   Obs.~1                      &          Obs.~2        & Obs.~3                  &    Obs.~4        &    Obs.~5              &       Obs.~6     \\
 \hline
                                &                               &                        &  {\rm A}    &                       &                        &                     \\
 \hline
   ${\rm N}_{\rm H}$\,$({\times}10^{22})\,({\rm cm}^{-2})$      &   $0.113{\pm}0.002$    &       $=$               &      $=$              &     $=$                &        $=$             &        $=$              \\
   ${\rm kT}_{1}$\,(keV)        &  $0.195{\pm}0.003$          &   $=$                  &       $=$               &      $=$              &     $=$                &        $=$              \\
   ${\rm kT}_{2}$\,(keV)        &  $1.000{\pm}0.011$          &   $=$                  &       $=$               &      $=$              &     $=$                &        $=$              \\
   ${\rm F}_{\rm X,D}$\,$^3$    &  $0.39{\pm}0.02$           &    $=$                 &  $=$                    &  $=$                  &  $=$                   &  $=$                  \\
   ${\rm kT}_{\rm in}$\,(keV)   &   $0.156{\pm}0.002$           &  $0.157{\pm}0.004$     &  $0.161{\pm}0.003$      &  $0.160{\pm}0.003$    &  $0.160{\pm}0.003$     &  $0.160{\pm}0.004$          \\
   ${\rm N}_{\rm disc}$         &   $221{\pm}21$                &  $179{\pm}20$     &  $170{\pm}21$           &  $148{\pm}21$         &  $167{\pm}20$          &  $136{\pm}20$               \\
   ${\Gamma}$                   &   $2.00{\pm}0.02$             &  $1.89{\pm}0.18$       &  $2.08{\pm}0.10$        &  $2.27{\pm}0.09$      &  $2.09{\pm}0.11$       &  $2.32{\pm}0.10$           \\
   ${\rm E}_{\rm c}$\,(keV)     &   $4.4{\pm}0.8$               &  $4.6{\pm}1.2$         &  $5.6{\pm}1.0$          &  $8.5{\pm}1.9$        &  $5.7{\pm}1.2$         &  $16{\pm}6$          \\
   ${\rm F}_{\rm disc}$\,$^3$   & $1.33{\pm}0.05$ & $1.13{\pm}0.04$ & $1.21{\pm}0.05$ & $1.01{\pm}0.04$ & $1.16{\pm}0.05$  &  $0.93{\pm}0.04$                   \\
   ${\rm F}_{\rm disc}(0.3-1\,{\rm keV})$\,$^3$   & $1.26{\pm}0.05$ & $1.07{\pm}0.04$ & $1.14{\pm}0.05$ & $0.95{\pm}0.04$ & $1.10{\pm}0.04$  &  $0.88{\pm}0.03$                   \\
   ${\rm F}_{\rm disc}(1-10\,{\rm keV})$\,$^3$   & $0.065{\pm}0.003$ & $0.058{\pm}0.002$ & $0.068{\pm}0.003$ & $0.055{\pm}0.002$ & $0.064{\pm}0.003$  &  $0.051{\pm}0.002$                   \\
   ${\rm L}_{\rm disc}(0.3-10\,{\rm keV})$\,$^{6}$\,$({\rm erg}\,{\rm s}^{-1})$ &   $(3.70{\pm}0.14){\times}10^{39}$   &  $(3.14{\pm}0.11){\times}10^{39}$  &  $(3.37{\pm}0.14){\times}10^{39}$ & $(2.81{\pm}0.11){\times}10^{39}$ & $(3.22{\pm}0.14){\times}10^{39}$ & $(2.59{\pm}0.11){\times}10^{39}$ \\
   ${\rm F}_{\rm pow}$\,$^3$    & $1.45{\pm}0.06$ & $1.40{\pm}0.06$ & $2.03{\pm}0.08$ & $2.13{\pm}0.08$ &  $1.78{\pm}0.07$ &  $1.96{\pm}0.08$       \\
   ${\rm F}_{\rm pow}(0.3-1\,{\rm keV})$\,$^3$    & $0.72{\pm}0.03$ & $0.64{\pm}0.03$ & $1.01{\pm}0.04$ & $1.14{\pm}0.05$ &  $0.89{\pm}0.04$ &  $1.03{\pm}0.04$       \\
   ${\rm F}_{\rm pow}(1-10\,{\rm keV})$\,$^3$    & $0.73{\pm}0.03$ & $0.76{\pm}0.03$ & $1.02{\pm}0.04$ & $0.99{\pm}0.04$ &  $0.89{\pm}0.04$ &  $0.93{\pm}0.04$       \\
   ${\rm F}_{\rm X,S}(0.3-1\,{\rm keV})$\,$^{3}$    & $1.98{\pm}0.08$ & $1.71{\pm}0.07$ & $2.15{\pm}0.09$ & $2.10{\pm}0.08$ & $2.00{\pm}0.08$  &  $1.90{\pm}0.08$            \\
   ${\rm F}_{\rm X,S}(1-10\,{\rm keV})$\,$^{3}$    & $0.80{\pm}0.03$ & $0.82{\pm}0.03$ & $1.09{\pm}0.04$ & $1.04{\pm}0.04$ & $0.95{\pm}0.04$  &  $0.98{\pm}0.04$            \\
   ${\rm F}_{\rm X,T}$\,$^{3}$    & $3.17{\pm}0.13$ & $2.92{\pm}0.12$ & $3.63{\pm}0.14$ & $3.53{\pm}0.14$ & $3.33{\pm}0.13$  &  $3.28{\pm}0.13$            \\
   ${\chi}^{2}/{\nu}=681/688$         &                               &                        &                         &                       &                        &                             \\
 \hline
                                &                               &                        &  {\rm B}    &                       &                        &                     \\
 \hline
   ${\rm N}_{\rm H}$\,$({\times}10^{22})\,({\rm cm}^{-2})$      &  $0.113{\pm}0.002$     &     $=$                 &     $=$               &      $=$               &    $=$                 &                         \\
   ${\rm kT}_{1}$\,(keV)          &  $0.195{\pm}0.004$           &  $=$                   &    $=$                  &    $=$                &     $=$                &   $=$                   \\  
   ${\rm kT}_{2}$\,(keV)          &  $1.000{\pm}0.011$           &  $=$                   &    $=$                  &    $=$                &     $=$                &   $=$                   \\  
   ${\rm F}_{\rm X,D}$\,$^3$     &  $0.39{\pm}0.02$          &   $=$                 &  $=$                    &  $=$                  &  $=$                   &  $=$                  \\
   ${\rm kT}_{\rm max}$\,(keV)    &  $0.148{\pm}0.002$            &  $0.150{\pm}0.004$     &  $0.153{\pm}0.003$      &  $0.152{\pm}0.003$    &  $0.152{\pm}0.003$     &  $0.152{\pm}0.003$          \\
   ${\rm N}_{\rm disc}$           &  $(4.0{\pm}0.4){\times}10^{-3}$ &  $(3.2{\pm}0.4){\times}10^{-3}$   &  $(3.1{\pm}0.4){\times}10^{-3}$    &  $(2.7{\pm}0.4){\times}10^{-3}$  &  $(3.0{\pm}0.4){\times}10^{-3}$  &  $(2.4{\pm}0.4){\times}10^{-3}$   \\
   ${\Gamma}$                   &  $2.01{\pm}0.02$              &  $1.90{\pm}0.18$       &  $2.08{\pm}0.10$        &  $2.28{\pm}0.09$      &  $2.10{\pm}0.11$       &  $2.32{\pm}0.10$           \\
   ${\rm E}_{\rm c}$\,(keV)     &  $4.5{\pm}0.7$                & $4.6{\pm}1.2$          &  $5.7{\pm}1.0$        & $8.6{\pm}2.0$         &  $5.8{\pm}1.2$         &  $16{\pm}7$               \\     
   ${\rm F}_{\rm disc}$\,$^3$   &  $1.32{\pm}0.05$   & $1.12{\pm}0.04$         &  $1.21{\pm}0.05$         &  $1.01{\pm}0.04$     &  $1.16{\pm}0.05$      &  $0.93{\pm}0.04$               \\
   ${\rm F}_{\rm disc}(0.3-1\,{\rm keV})$\,$^3$   & $1.26{\pm}0.05$ & $1.07{\pm}0.04$ & $1.14{\pm}0.05$ & $0.95{\pm}0.04$ & $1.10{\pm}0.04$  &  $0.88{\pm}0.03$                   \\
   ${\rm F}_{\rm disc}(1-10\,{\rm keV})$\,$^3$   & $0.064{\pm}0.003$ & $0.057{\pm}0.002$ & $0.067{\pm}0.003$ & $0.054{\pm}0.002$ & $0.063{\pm}0.002$  &  $0.051{\pm}0.002$                   \\
   ${\rm L}_{\rm disc}(0.3-10\,{\rm keV})$\,$^{6}$\,$({\rm erg}\,{\rm s}^{-1})$ &   $(3.67{\pm}0.14){\times}10^{39}$   &  $(3.12{\pm}0.11){\times}10^{39}$  &  $(3.37{\pm}0.14){\times}10^{39}$ & $(2.81{\pm}0.11){\times}10^{39}$ & $(3.23{\pm}0.14){\times}10^{39}$ & $(2.59{\pm}0.11){\times}10^{39}$ \\
   ${\rm F}_{\rm pow}$\,$^3$    &  $1.46{\pm}0.06$   & $1.41{\pm}0.06$         &  $2.04{\pm}0.08$         &  $2.14{\pm}0.08$     &  $1.79{\pm}0.07$      &  $1.97{\pm}0.08$       \\
   ${\rm F}_{\rm pow}(0.3-1\,{\rm keV})$\,$^3$    & $0.73{\pm}0.03$ & $0.65{\pm}0.03$ & $1.02{\pm}0.04$ & $1.15{\pm}0.05$ &  $0.90{\pm}0.04$ &  $1.03{\pm}0.04$       \\
   ${\rm F}_{\rm pow}(1-10\,{\rm keV})$\,$^3$    & $0.73{\pm}0.03$ & $0.76{\pm}0.03$ & $1.02{\pm}0.04$ & $0.99{\pm}0.04$ &  $0.89{\pm}0.04$ &  $0.93{\pm}0.04$       \\
   ${\rm F}_{\rm X,S}(0.3-1\,{\rm keV})$\,$^{3}$    & $1.98{\pm}0.08$ & $1.71{\pm}0.07$ & $2.16{\pm}0.09$ & $2.10{\pm}0.08$ & $2.00{\pm}0.08$  &  $1.90{\pm}0.08$            \\
   ${\rm F}_{\rm X,S}(1-10\,{\rm keV})$\,$^{3}$    & $0.80{\pm}0.03$ & $0.82{\pm}0.03$ & $1.09{\pm}0.04$ & $1.04{\pm}0.04$ & $0.95{\pm}0.04$  &  $0.98{\pm}0.04$            \\
   ${\rm F}_{\rm X,T}$\,$^{3}$    & $3.17{\pm}0.13$ & $2.93{\pm}0.12$ & $3.63{\pm}0.14$ & $3.54{\pm}0.14$ & $3.34{\pm}0.13$  &  $3.29{\pm}0.13$            \\
   ${\chi}^{2}/{\nu}=681/688$         &                               &                        &                         &                       &                        &                             \\
\hline
\end{tabular}
\footnotetext{$^1$ The spectral models used are: A) {\tt tbabs(apec+apec+diskbb+cutoffpl)}, B) {\tt tbabs(apec+apec+diskpn+cutoffpl)}, C) {\tt tbabs(apec+apec+diskbb+comptt)} and D) {\tt tbabs(apec+apec+diskpn+comptt)}.}
\footnotetext{$^2$ Errors are 68\% confidence errors. }
\footnotetext{$^3$ Unabsorbed flux in the 0.3--10\,keV energy range, in units of ${\times}10^{-12}{\rm erg}{\rm s}^{-1}{\rm cm}^{-2}$. }
\footnotetext{Description of the parameters:\\
1) Total column density ${\rm N}_{\rm H}$; 2) temperatures from the emission components describing the diffuse emission from the galaxy (${\rm kT}_{1}$ and ${\rm kT}_{2}$);
3) unabsorbed flux from the diffuse emission of the galaxy and from the source (${\rm F}_{\rm X,D}$, ${\rm F}_{\rm X,S}$); 4) temperature from the inner accretion disc (${\rm kT}_{\rm in}$ and ${\rm kT}_{\rm max}$ for the {\tt diskbb} and {\tt diskpn} components, respectively) 
and normalization from the disc component (${\rm N}_{\rm disc}$); 5) temperature of the electrons and opacity of the corona (${\rm kT}_{\rm e}$,${\tau}$) from the {\tt compTT} model component;
6) power-law photon index (${\Gamma}$) and e-folding energy of the exponential rolloff (${\rm E}_{\rm c}$) for the {\tt cutoffpl} model component; 7) unabsorbed flux 
of the disc emission component (${\rm F}_{\rm disc}$) and corresponding luminosity (${\rm L}_{\rm disc}$; assuming a distance of 4.8\,Mpc) in the 0.3-10\,keV energy range; 8) unabsorbed flux from the high-energy emission 
component (${\rm F}_{\rm pow}$ and ${\rm F}_{\rm compTT}$; for the 
{\tt cutoffpl} or {\tt compTT} model components, respectively) in the 0.3-10\,keV energy range; and 9) unabsorbed total flux (source plus diffuse emission from the galaxy) in the 0.3-10\,keV energy range (${\rm F}_{\rm X,T}$).}
\end{minipage}
\end{table*}

\begin{table*}
 \centering
   \addtocounter{table}{-1}
 \begin{minipage}{160mm}
  \caption{(Continued.)}
  \begin{tabular}{@{}lcccccc@{}}
  \hline
   Spectral parameter~$^1$$^,$$^2$ &   Obs.~1                      &          Obs.~2        & Obs.~3                  &    Obs.~4        &    Obs.~5              &       Obs.~6     \\
 \hline
                                &                               &                        &  {\rm C}    &                       &                        &                     \\
 \hline
   ${\rm N}_{\rm H}$\,$({\times}10^{22})\,({\rm cm}^{-2})$      &  $0.108{\pm}0.002$     &    $=$                  &   $=$                 &     $=$                &  $=$                   &    $=$                  \\
   ${\rm kT}_{1}$\,(keV)          &   $0.195{\pm}0.003$          &    $=$                 &    $=$                  &    $=$                &   $=$                  &  $=$                    \\
   ${\rm kT}_{2}$\,(keV)          &   $1.001{\pm}0.011$           &    $=$                 &    $=$                  &    $=$                &   $=$                  &  $=$                    \\
   ${\rm F}_{\rm X,D}$\,$^3$    &  $0.37{\pm}0.02$            & $=$                 &  $=$                    &  $=$                  &  $=$                   &  $=$                  \\
   ${\rm kT}_{\rm in}$\,(keV)   &  $0.140{\pm}0.002$            & $0.140{\pm}0.004$      & $0.140{\pm}0.003$       &  $0.137{\pm}0.004$    &  $0.140{\pm}0.004$     &  $0.138{\pm}0.004$          \\
   ${\rm N}_{\rm disc}$         &  $388{\pm}9$                 & $330{\pm}40$            & $390{\pm}30$            &  $410{\pm}40$         &  $360{\pm}30$          &  $360{\pm}30$                  \\
   ${\rm kT}_{\rm e}$\,(keV)    &  $1.0_{-1.0}^{+0.1}$         & $1.0_{-1.0}^{+0.13}$   & $1.0_{-1.0}^{+0.12}$    &  $1.0_{-1.0}^{+0.3}$  &  $1.0_{-1.0}^{+0.3}$   &  $4.2{\pm}1.6$          \\
   ${\tau}$                     &  $5.07{\pm}0.11$              & $5.2{\pm}0.2$         & $5.1{\pm}0.2$         &  $4.8{\pm}0.5$       &  $4.8{\pm}0.5$        &  $3.5{\pm}0.9$                \\
   ${\rm F}_{\rm disc}$\,$^3$   & $1.37{\pm}0.05$    &  $1.15{\pm}0.05$        &  $1.34{\pm}0.05$         &  $1.27{\pm}0.05$     &  $1.28{\pm}0.05$      &   $1.20{\pm}0.05$              \\
   ${\rm F}_{\rm disc}(0.3-1\,{\rm keV})$\,$^3$   & $1.32{\pm}0.05$ & $1.11{\pm}0.04$ & $1.30{\pm}0.05$ & $1.23{\pm}0.05$ & $1.24{\pm}0.05$  &  $1.16{\pm}0.05$                   \\
   ${\rm F}_{\rm disc}(1-10\,{\rm keV})$\,$^3$   & $0.044{\pm}0.002$ & $0.037{\pm}0.002$ & $0.043{\pm}0.002$ & $0.037{\pm}0.002$ & $0.041{\pm}0.002$  &  $0.037{\pm}0.002$                   \\
   ${\rm L}_{\rm disc}(0.3-10\,{\rm keV})$\,$^{6}$\,$({\rm erg}\,{\rm s}^{-1})$ &   $(3.81{\pm}0.14){\times}10^{39}$   &  $(3.20{\pm}0.14){\times}10^{39}$  &  $(3.73{\pm}0.14){\times}10^{39}$ & $(3.53{\pm}0.14){\times}10^{39}$ & $(3.56{\pm}0.14){\times}10^{39}$ & $(3.34{\pm}0.14){\times}10^{39}$ \\
   ${\rm F}_{\rm compTT}$\,$^3$    & $1.33{\pm}0.05$    &  $1.36{\pm}0.05$        &  $1.80{\pm}0.07$         &  $1.75{\pm}0.07$     &  $1.58{\pm}0.06$      &   $1.59{\pm}0.06$         \\
   ${\rm F}_{\rm compTT}(0.3-1\,{\rm keV})$\,$^3$    & $0.58{\pm}0.02$ & $0.55{\pm}0.02$ & $0.76{\pm}0.03$ & $0.76{\pm}0.03$ &  $0.67{\pm}0.03$ &  $0.65{\pm}0.03$       \\
   ${\rm F}_{\rm compTT}(1-10\,{\rm keV})$\,$^3$    & $0.75{\pm}0.03$ & $0.81{\pm}0.03$ & $1.05{\pm}0.04$ & $1.00{\pm}0.04$ &  $0.91{\pm}0.04$ &  $0.94{\pm}0.04$       \\
   ${\rm F}_{\rm X,S}(0.3-1\,{\rm keV})$\,$^{3}$    & $1.90{\pm}0.08$ & $1.65{\pm}0.07$ & $2.06{\pm}0.08$ & $2.00{\pm}0.08$ & $1.91{\pm}0.08$  &  $1.81{\pm}0.07$            \\
   ${\rm F}_{\rm X,S}(1-10\,{\rm keV})$\,$^{3}$    & $0.80{\pm}0.03$ & $0.85{\pm}0.03$ & $1.09{\pm}0.04$ & $1.03{\pm}0.04$ & $0.95{\pm}0.04$  &  $0.97{\pm}0.04$            \\
   ${\rm F}_{\rm X,T}$\,$^{3}$    & $3.07{\pm}0.12$ & $2.87{\pm}0.11$ & $3.52{\pm}0.14$ & $3.39{\pm}0.13$ & $3.22{\pm}0.13$  &  $3.16{\pm}0.13$            \\
   ${\chi}^{2}/{\nu}=698/688$         &                               &                        &                         &                       &                        &                             \\
 \hline
                                &                               &                        &  {\rm D}    &                       &                        &                     \\
 \hline
   ${\rm N}_{\rm H}$\,$({\times}10^{22})\,({\rm cm}^{-2})$      &  $0.108{\pm}0.004$     &  $=$                    &  $=$                  &    $=$                 &     $=$                &                         \\
   ${\rm kT}_{1}$\,(keV)          &   $0.194{\pm}0.005$           &    $=$                 &   $=$                   &  $=$                  &    $=$                 &   $=$                         \\
   ${\rm kT}_{2}$\,(keV)          &   $1.003{\pm}0.011$           &    $=$                 &   $=$                   &  $=$                  &    $=$                 &   $=$                   \\
   ${\rm F}_{\rm X,D}$\,$^3$    &  $0.36{\pm}0.02$              & $=$                 &  $=$                    &  $=$                  &  $=$                   &  $=$                  \\
   ${\rm kT}_{\rm max}$\,(keV)  & $0.134{\pm}0.003$             & $0.134{\pm}0.004$         & $0.134{\pm}0.003$     & $0.131{\pm}0.004$        & $0.133{\pm}0.002$         & $0.133{\pm}0.003$           \\
   ${\rm N}_{\rm disc}$         & $(6.6{\pm}0.4){\times}10^{-3}$ &  $(5.6{\pm}0.6){\times}10^{-3}$  & $(6.6{\pm}0.6){\times}10^{-3}$  &  $(6.8{\pm}0.7){\times}10^{-3}$ & $(6.2{\pm}0.5){\times}10^{-3}$ & $(6.1{\pm}0.6){\times}10^{-3}$                         \\
   ${\rm kT}_{\rm e}$\,(keV)    & $1.0_{-1.0}^{+0.2}$          & $1.0_{-1.0}^{+0.2}$    & $1.0_{-1.0}^{+0.2}$     & $1.0_{-1.0}^{+0.3}$        & $1.0_{-1.0}^{+0.3}$    & $4.6{\pm}1.9$                        \\  
   ${\tau}$                     & $5.0{\pm}0.2$               & $5.2{\pm}0.2$         & $5.1{\pm}0.2$         & $4.9{\pm}0.4$       & $4.8{\pm}0.4$          & $3.4{\pm}1.0$           \\ 
   ${\rm F}_{\rm disc}$\,$^3$   &   $1.32{\pm}0.05$             & $1.11{\pm}0.04$        &  $1.29{\pm}0.05$        & $1.21{\pm}0.05$       & $1.22{\pm}0.05$        & $1.14{\pm}0.05$ \\
   ${\rm F}_{\rm disc}(0.3-1\,{\rm keV})$\,$^3$   & $1.28{\pm}0.05$ & $1.07{\pm}0.04$ & $1.25{\pm}0.05$ & $1.18{\pm}0.05$ & $1.18{\pm}0.05$  &  $1.11{\pm}0.04$                   \\
   ${\rm F}_{\rm disc}(1-10\,{\rm keV})$\,$^3$   & $0.042{\pm}0.002$ & $0.036{\pm}0.002$ & $0.041{\pm}0.002$ & $0.036{\pm}0.002$ & $0.039{\pm}0.002$  &  $0.035{\pm}0.002$                   \\
   ${\rm L}_{\rm disc}(0.3-10\,{\rm keV})$\,$^{6}$\,$({\rm erg}\,{\rm s}^{-1})$ &   $(3.67{\pm}0.14){\times}10^{39}$   &  $(3.09{\pm}0.11){\times}10^{39}$  &  $(3.59{\pm}0.14){\times}10^{39}$ & $(3.37{\pm}0.14){\times}10^{39}$ & $(3.39{\pm}0.14){\times}10^{39}$ & $(3.17{\pm}0.14){\times}10^{39}$ \\
   ${\rm F}_{\rm compTT}$\,$^3$    &   $1.36{\pm}0.05$             & $1.38{\pm}0.05$        &  $1.85{\pm}0.07$        & $1.80{\pm}0.07$       & $1.62{\pm}0.06$        & $1.63{\pm}0.07$ \\
   ${\rm F}_{\rm compTT}(0.3-1\,{\rm keV})$\,$^3$    & $0.61{\pm}0.02$ & $0.57{\pm}0.02$ & $0.80{\pm}0.03$ & $0.80{\pm}0.03$ &  $0.71{\pm}0.03$ &  $0.69{\pm}0.03$       \\
   ${\rm F}_{\rm compTT}(1-10\,{\rm keV})$\,$^3$    & $0.75{\pm}0.03$ & $0.81{\pm}0.03$ & $1.05{\pm}0.04$ & $1.00{\pm}0.04$ &  $0.91{\pm}0.04$ &  $0.94{\pm}0.04$       \\
   ${\rm F}_{\rm X,S}(0.3-1\,{\rm keV})$\,$^{3}$    & $1.89{\pm}0.08$ & $1.64{\pm}0.07$ & $2.04{\pm}0.08$ & $1.98{\pm}0.08$ & $1.89{\pm}0.08$  &  $1.80{\pm}0.07$            \\
   ${\rm F}_{\rm X,S}(1-10\,{\rm keV})$\,$^{3}$    & $0.80{\pm}0.03$ & $0.85{\pm}0.03$ & $1.09{\pm}0.04$ & $1.03{\pm}0.04$ & $0.95{\pm}0.04$  &  $0.98{\pm}0.04$            \\
   ${\rm F}_{\rm X,T}$\,$^{3}$    & $3.04{\pm}0.12$ & $2.85{\pm}0.11$ & $3.49{\pm}0.14$ & $3.37{\pm}0.13$ & $3.20{\pm}0.13$  &  $3.13{\pm}0.12$            \\
   ${\chi}^{2}/{\nu}=697/688$         &                               &                        &                         &                       &                        &                             \\
\hline
\end{tabular}
\end{minipage}
\end{table*}

\subsection{Results}

\begin{figure}
\centering
 \includegraphics[bb= 0 0 612 792,width=7.0cm,angle=270,clip]{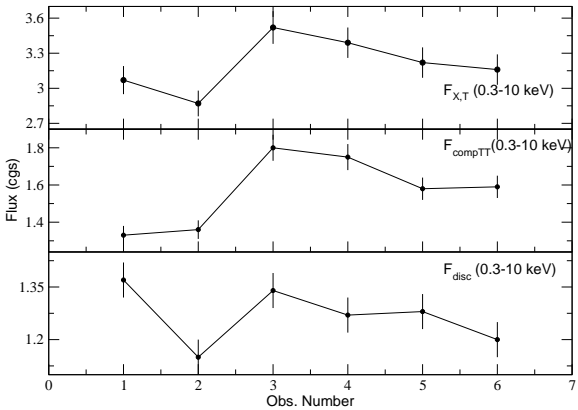}
 \caption{Unabsorbed fluxes (in the 0.3-10\,keV energy range) versus the observation number. Total (with emission from the diffuse component), high-energy and disc component unabsorbed fluxes (top, middle and bottom, respectively).
 Values are reported in Tab.~\ref{table_spe} and errors are $68\%$ confidence. Spectral model C has been used in this plot to derive the fluxes.}
 \label{plot_lc}
\end{figure}

\begin{figure}
\centering
 \includegraphics[bb=0 0 612 792,width=7.0cm,angle=270,clip]{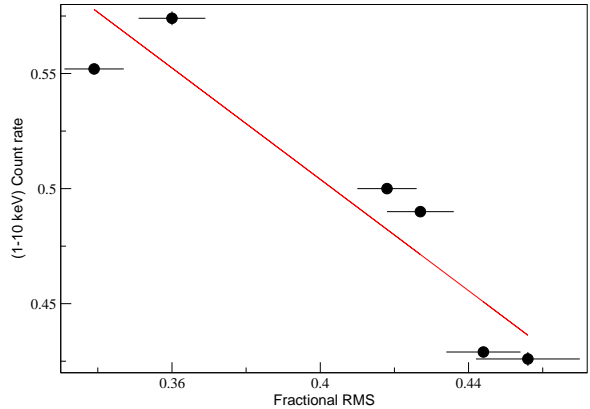}
 \caption{Count rate (in the 1-10\,keV energy range) versus the fractional rms of the variability in the 1-10\,keV energy range and 0.0001-0.19\,Hz frequency range. Errors are $68\%$ confidence. Poissonian errors
on the count rate are smaller than the symbol.  }
 \label{plot_anticorr}
\end{figure}

In Fig.~\ref{plot_lc} we show the light curve from the total flux (i.e. source plus emission from the diffuse component) and from the source, separated in its different 
components (disc and high-energy component -- the last also called Comptonization component). Focusing 
on the light curve behaviour from the source we see that the disc component shows no significant variations (within $2{\sigma}$ errors) in flux. On 
the other hand we see significant variations (at the $3{\sigma}$ level with respect to the mean value) in the flux from the high-energy component. These variations are very similar
to the variations from the total flux. Therefore, it looks like
the variations of the total flux are due to variations of the high-energy component. Additionally, the broad-band noise in the 1-10\,keV shows significant 
variations (at the $3{\sigma}$ level with respect to the mean value). These two quantities (broad-band noise and total/high-energy component flux) appear to be anti-correlated, with a linear correlation
coefficient of -0.87 (using the least-squares procedure). This corresponds to a chance probability of $0.6\%$. Also, the (1-10\,keV) count rate is anti-correlated with the fractional rms, with a linear correlation
coefficient of -0.93. This corresponds to a chance probability of $0.1\%$ (see Fig.~\ref{plot_anticorr}). Therefore, the anti-correlation is not connected to the specific 
spectral model used.  

\section{Discussion}

In this paper we report on the timing and spectral analysis of the longest {\it XMM-Newton} observations of NGC~5408 X--1. 
The only significant ($>3{\sigma}$) spectral variation is the flux from the high-energy component (see Fig.~\ref{plot_lc}).
In the following we assume that the {\it soft excess} is physically a disc component, in order to compare with the case of BHBs.
The variations in the flux from the disc component are less than $2\,{\sigma}$ with respect to the mean value (where ${\sigma}{\approx}0.05{\times}10^{-12}\,{\rm erg}\,{\rm s}^{-1}$ is the mean error from the flux values). 
Regarding the fast time variability from NGC~5408 X--1, the fractional 
rms in the 0.3--1\,keV energy range does not significantly vary between the observations and is ${\approx}(12-15)\%$ in the $10^{-4}-0.19$\,Hz frequency range, in agreement with
previous studies \citep{heil10,middleton11}. On the contrary, 
the fractional rms in the 1--10\,keV energy range is high (in agreement with the previous studies) and variable between the observations (${\approx}30-50\%$). 
In the following our goal is to compare our findings with the case of BHBs.

Traditionally, the variability of BHBs has been associated to the high-energy emission component (e.g. \citealt{galeev79,poutanen99,churazov01,done07}). This is because, although strong variations
in the luminosity of the disc component are seen on the long time-scales corresponding to transitions between the soft and the hard spectral states, the disc-dominated soft
states show very little rapid variability (fractional rms${\approx}1\%$; e.g. \citealt{belloni05}). In the low/hard state of BHBs the understanding of the origin of the variability has been limited by
a lack of spectral coverage of the disc component, which emits at $<1$\,keV, and so is not covered by the band-pass of detectors such as the {\it Proportional Counter Array} (PCA) on the {\it RXTE} satellite (2-60\,keV). Nevertheless, 
a recent study of the X-ray variability in the low/hard state of BHBs has been performed with {\it XMM-Newton} \citep{wilkinson09}, who have found a disc component variability with a fractional rms of
${\approx}30-40\%$ for SWIFT~J1753.5--0127 and GX~339--4 over long time-scales (2.7-270\,s) when the source was in the low/hard state. From our work we see that in the case of NGC~5408 X--1 
the 0.3-1\,keV energy range is dominated by emission from the inner disc
(contributing $60\%$ of the total flux, with the high-energy component contributing $30\%$ only). We have found that the fractional variability from NGC~5408 X--1 in the 0.3-1\,keV 
energy range and $10^{-4}-0.19$\,Hz frequency range is of 
${\approx}15\%$. This is less than what has been found in the {\it XMM-Newton} band-pass for BHBs. The difference
might be due to the fact that both the disc and the power-law components contribute differently to the 0.3--1\,keV energy range in the case of NGC~5408 X--1 with respect to the case of BHBs, since the latter have hotter inner disc emission. 

In contrast, rms of several tens per cent is seen in the low/hard spectral state of BHBs
which are energetically dominated by the high-energy component emission in the {\it RXTE} band-pass (2-15\,keV). 
We have found that in the case of NGC~5408 X--1 the 1-10\,keV energy range is dominated by the high-energy emission component (contributing $90\%$ of the total flux) and that little emission comes from the accretion disc in the 
1-10\,keV energy range ($4\%$), similarly to what has been found in GX~339--4 in the 2--15\,keV energy range during the low/hard state. This fact means that we can only use our results from the study of the fast variability 
of NGC~5408 X--1 in the 1--10\,keV band for the comparison with the case of BHBs (i.e. measured with {\it RXTE}), since the contribution to the flux from the disc component are small in both cases.
It is in this energy range (i.e. 1-10\,keV) where we are measuring fractional rms variabilities
similar to the high-energy component during the low/hard state of GX~339--4 (i.e. $30-40\%$). 

From a detailed study of the variability of the BHB GX~339--4 with {\it RXTE} it has been seen that its variability decreases
as its total flux (and the contribution from the disc component flux) increases. This has also seen to occur in a wider sample of BHBs using {\it RXTE} data only \citep{heil12}.
We discovered that the the (1-10\,keV) count rate from NGC~5408 X--1 is anti-correlated with
the fractional rms over long time-scales (years), similarly to the behaviour of the BHB GX~339-4 during the bright stage of the low/hard state measured with the {\it RXTE} satellite (\citealt{munoz11}; hereafter called bright hard state). 
The relationship found by \citet{munoz11} for GX~339-4 is valid over time-scales of weeks to months, thus shorter than in our case. 
Also, the noise in the PDS of NGC~5408 X--1 is in a lower frequency interval (0.0001-0.19\,Hz) than GX~339-4 in the low/hard state (0.1-64\,Hz frequency range; \citealt{munoz11}).
The fact that the characteristic time-scales of variability appear at lower frequency in the case of NGC~5408 X--1 with respect to GX~339-4 might be
due to the higher mass of the BH and accretion rate in the former, as previously noticed to occur in a comparison in the case of BHBs with their higher mass analogs, i.e. the AGN 
\citep{mchardy06}.

Previous studies revealed that a linear absolute rms versus flux positive correlation occurs on short
time-scales (few ks) in NGC~5408 X--1 \citep{heil10}. Contrary to our work, this result was found on short timescales where the PDS is restricted to being
stationary. On the other hand, a recent study of the timing behaviour of BHBs \citep{heil12} but using {\it RXTE} data (${\rm E}{\ge}2$\,keV), 
has shown that the same rms-flux anti-correlation we have found here holds in the bright hard state of GX~339-4 (see their Fig.\,4-5). This is in agreement with the 
results reported earlier by \citet{munoz11}. Both studies and our work use the non-stationary rms-flux on long timescales which samples changes to the whole PDS.  

\subsection{On the Mass of the Black-Hole as found from previous studies}  \label{sec1}

As described in Sec.\,\ref{introd}, determining the mass from the BH in NGC~5408 X--1 has been the goal of several studies. However, there is still no consensus
on whether it is an IMBH or a stellar-mass BH. Previous estimates from the timing properties \citep{strohmayer09,dheeraj12} indicate a mass of ${\rm M}_{\rm BH}{\gtrsim}1\,000\,{\rm M}_{\odot}$, thus an IMBH,
but others \citep{middleton11} indicate a much smaller mass of ${\rm M}_{\rm BH}{\le}100\,{\rm M}_{\odot}$, thus a stellar-mass BH \footnote{It has to be noted here that in low metallicity environments
BHs with masses up to $80-100\,{\rm M}_{\rm BH}$ can still be formed through direct stellar-collapse (\citealt{zampieri09,belczynski10}) and this is why we are referring to them as {\it stellar-mass} BHs.}. In the 
first case considered the accretion rate is sub-Eddington, whilst the latter case indicates (near or) super-Eddington accretion. 

In the following we use the results obtained from the spectral analysis. The low inner disc temperatures found for some ULXs were interpreted as an evidence for the presence of IMBH \citep{miller03,miller04}. 
In the standard disc-black body model (i.e. Multi-Color Disc Blackbody or MCD; \citealt{makishima86,makishima00}), which is a very poor approximation of the real standard accretion disc theory \citep{frank02}, the bolometric luminosity 
from the accretion disc is calculated as:

\begin{equation} \label{eq1}
{\rm L}_{\rm bol}=4{\pi}({\rm R}_{\rm in}/{\zeta})^{2}{\sigma}({\rm T}_{\rm in}/{\kappa})^{4}
\end{equation}

\noindent Here ${\kappa}{\approx}1.7$ \citep{shimura95} is the ratio of the color temperature to the effective temperature, or ''spectral hardening
factor'', and ${\zeta}$ is a correction factor taking into account the fact that ${\rm T}_{\rm in}$ occurs at a radius somewhat larger than ${\rm R}_{\rm in}$
(\citealt{kubota98} give ${\zeta}=0.412$). The behaviour explained above is a viable explanation of the temperature-luminosity behaviour of 
most BHBs (see \citealt{done07} and references therein). However, a recent spectral study of the spectral variability 
from a sample of ULXs, including NGC~5408 X--1 \citep{kajava09}, has shown that the {\it soft excess} (i.e. the disc component fitted in the spectra) from NGC~5408 X--1 
does not follow Eq.~\ref{eq1} but ${\rm L}_{\rm bol}{\propto}{\rm T}_{\rm in}^{-3.5}$. This in contrast to what is found for many BHBs and might indicate that
the standard accretion disc theory is not is not a proper interpretation in the case of NGC~5408 X--1. This implies that the hypothesis on which 
the IMBH idea is relying (i.e. standard accretion disc theory and the presence of a cold disc) are not valid and 
it might indicate that the BH in NGC~5408 X--1 is not an IMBH.

Therefore, other inner disc configurations might be possible in the case of NGC~5408 X--1.
The large apparent luminosities in the case of ULXs can be explained by supercritical accretion (exceeding the Eddington luminosity) onto a stellar-mass BH. In a recent study \citep{poutanen07} it has been proposed 
that at high accretion rates an outflow forms within the so-called spherization radius. For a face-on observer the luminosity is high because of geometrical beaming \citep{king01}. Such an observer has a direct view of the
inner hot accretion disc, which has a peak temperature ${\rm T}_{\rm max}{\approx}1$\,keV in stellar-mass BHs. In this model the {\it soft excess} corresponds to the emission from the spherization radius.
Therefore, having a stellar-mass BH implies the presence of a much hotter inner accretion disc (i.e. with temperatures higher than the {\it soft excess}), that would be 
observed in the spectrum if the inner disc inclination is low. This could be the responsible
for the ``Comptonization'' component seen in the spectra from NGC~5408 X--1, that could be emission from the inner disc instead. Such a super-Eddington flow implies much lower values for the mass of the BH, 
i.e. ${\rm M}{\approx}10\,{\rm M}_{\odot}$, accreting at mildly super-Eddington rates (${\dot M}/{\dot M}_{\rm EDD}{\approx}10$).
An alternative scenario has been proposed \citep{stobbart06,roberts07,soria07,gladstone09}, in which the presence of a cold and optically thick corona is obscuring the inner region of the disc and in this case the MCD parameters
cannot provide reliable estimates of the black hole mass. Again this scenario is compatible with a stellar-mass BH, i.e. ${\rm M}{\approx}10-100\,{\rm M}_{\odot}$. These models are consistent with the idea that
there is a relatively stable component at soft energies diluting the variability as might be expected by a thermal disc (see \citealt{churazov01}) or optically thick photospheric component
\citep{zdziarski09,zdziarski10}.  

\subsection{The accretion state}

In this paper we have shown that the timing properties of NGC~5408 X--1 over long time-scales (years) show that the variability properties, in 
particular the rms, change and resemble 
those from the BHB GX~339--4 during the bright phase of the hard state. These properties are the presence of a high
broad-band timing noise and an anti-correlation of the fractional rms versus the high-energy count rate.

In the case of BHBs the hard state is not only seen during low accretion rates. Indeed, high accretion levels have been reached in a few cases. In the case of GS 2023$+$338 and Cyg~X--1; 
\citep{tanaka95,stern01} quasi-Eddington luminosities have been observed during the hard state. NGC~5408 X--1 has similar properties to what is found in the BHB GX~339--4 during its bright hard 
state. Additionally, in the hard state of GX~339--4 the major spectral change is in the flux from the high-energy component
(see Fig.~6 of \citealt{motta09}). This behaviour is again similar to what is seen in the case of NGC~5408 X--1 (Fig.~\ref{plot_lc}).

Nevertheless, an identification of the accretion state with the low/hard state seen in BHBs can not be made. The spectral properties of NGC~5408 X--1 are like ULXs in the so called 
``power-law state'' \citep{makishima07,soria11}. As mentioned in Sec.~\ref{introd}, during the ``power-law state'' the
spectra of ULXs show a power-law spectral shape in the 3-8\,keV spectral range, together with a high-energy
turn-over at 6-7\,keV, and a {\it soft excess} at low energies (e.g. \citealt{kaaret06}). There might be a corona in the case of NGC~5408 X--1 (and other ULXs) that is much colder and 
thicker than in the case of BHBs. Alternatively, there could be an intense accretion wind that is producing both the spectrum shape and the variability observed. The 
different properties could be due simply because of a much higher Eddington ratio in the case of NGC~5408 X--1. 

The absolute rms of NGC~5408 X-1 in our observations is compatible with being constant within $1{\sigma}$ errors. 
This is not compatible with a low/hard state-like behaviour in BHBs (and GX~339--4 in particular), where the absolute rms increases with count rate \citep{munoz11}. Alternatively,
this behaviour is similar to what has been seen in GX~339--4 during the Hard Intermediate state, in particular at the point where its fractional variability is of $20\%$ and where the presence
of an accretion disc is detected for the first time \citep{munoz11}. In this state the fractional rms decreases as the count rate increases, as we observe to occur in NGC~5408 X--1. The fact that 
the fractional rms is of the order of ${\approx}20\%$ in the case of
GX~339--4 whilst in the case of NGC~5408 X--1 is of the order of ${\approx}40\%$ is not problematic, since it can be explained by the presence of a much colder accretion disc 
(see Sec.~\ref{sec1}) for the ULX. This decreases the contribution of the accretion disc component in the 1-10\,keV energy range and diminishes the effect 
of suppression of variability in this energy range for the ULX.

\section*{Acknowledgments}

We thank the anonymous referee, S. E. Motta, L. Zampieri, R. Soria and J.~C. Gladstone for discussions and insights. 
This work is based on observations made with {\it XMM-Newton}, an ESA science mission with instruments and contributions directly
funded by ESA member states and the USA (NASA). MCG acknowledges support from INAF through a 2010 postdoctoral fellowship. 
TB and AW acknowledge support from grant ASI-INAF I/009/10/. The research leading to these results has received
funding from the European Community’s Seventh Framework Programme (FP7/2007-2013) under grant agreement number ITN 215212 Black Hole Universe.
This research has made use of the General High-energy Aperiodic Timing Software (GHATS) package developed by T.M. Belloni at INAF - Osservatorio Astronomico di Brera.

\end{document}